\documentclass[superscriptaddress,aps,prl,twocolumn]{revtex4-2}
\usepackage{graphicx}
\usepackage{epstopdf}
\usepackage{bm,amsmath,amssymb} % for math
\usepackage{color, soul} % for highlight
\usepackage{dcolumn}   % needed for some tables
\usepackage{multirow}
\newcommand{\ie}{\textit{i}.\textit{e}.}
\newcommand{\eg}{{\em e.\,g.}}
\begin{document}
\title{Three Dimensional Activity Volcano Plot under External Electric Field }%

\author{Changming Ke}
\affiliation{Key Laboratory for Quantum Materials of Zhejiang Province, School of Science, Westlake University, Hangzhou, Zhejiang 310024, China}
\affiliation{Institute of Natural Sciences, Westlake Institute for Advanced Study, Hangzhou, Zhejiang 310024, China}
\author{Zijing Lin}
\affiliation{Hefei National Laboratory for Physical Sciences at Microscales, CAS Key Laboratory of Strongly-Coupled Quantum Matter Physics, Department of Physics, University of Science and Technology of China, Hefei 230026, China}
\author{Shi Liu}
\email{liushi@westlake.edu.cn}
\affiliation{Key Laboratory for Quantum Materials of Zhejiang Province, School of Science, Westlake University, Hangzhou, Zhejiang 310024, China}
\affiliation{Institute of Natural Sciences, Westlake Institute for Advanced Study, Hangzhou, Zhejiang 310024, China}

%\date{\today}

\begin{abstract}{ 
An external electric field (EEF) can impact a broad range of catalytic processes beyond redox systems. Computational design of catalysts under EEFs targeting specific operation conditions essentially requires accurate predictions of the response of a complex physicochemical system to collective parameters such as EEF strength/direction and temperature. Here, we develop a first-principles-based multiscale approach that enables efficient EEF-dependent kinetic modeling of heterogeneous catalysis. Taking steam reforming of methanol as an example, we find that the methanol conversion rate exhibits strong nonlinear response to temperature and EEF, and the optimal field line and constant carbon concentration line defined in the temperature--EEF parameter space serve as powerful metrics for catalyst design. Assisted with a deep neural network,  we establish a three-dimensional activity volcano plot under EEFs for thousands of metallic alloys.}

\end{abstract}

\maketitle
\newpage

%\section{Introduction}
Much like Earth's climate and spin glass, heterogeneous catalysis is an archetypal complex system and plays a pivotal role in our society. Electric field (EF) can manipulate the chemical reactivity and selectivity by tuning the relative stability of polar or ionic reactants and transition states in chemical reactions~\cite{Shaik16p1091,Shaik18p5125,Che18p5153,Ciampi18p5146,Shaik20p12551,Stuyver20pe1438}, opening up exciting opportunities to control catalytic reactions in an on-demand manner using EFs, hereinafter referred to as electrostatic catalysis. To take full advantage of electrostatic catalysis, a fundamental challenge is to predict the response of heterogeneous catalysis, a complex physicochemical system often involving a network of chemical reactions and physical processes, to  collective reaction parameters, specifically field strength and temperature.

The presence of an EF under realistic reaction conditions can have both intrinsic and extrinsic origins~\cite{Aragones16p88,Shaik18p5125,Ciampi18p5146,Stuyver20pe1438}. It is more convenient in practice to directly apply an external EF (EEF) to a reactor as the strength and direction of the field is readily tunable.  To apply EEFs in large-scale reactions, several reactor systems have been developed~\cite{Oshima13p3003,Oshima14p27,Sekine09p183}, \ie, probe-bed-probe reactors that have a conductive catalyst bed placed in the gap between two external probes and continuous-circuit reactors in which the conductive catalyst bed is integrated as an electrical circuit. The catalyst bed often consists of catalytically active metal nanoparticles that can generate very high surface EFs in the presence of EEFs due to the curvature-induced field enhancement effect~\cite{Gray20p125640}.
Additionally, surface EFs are always perpendicular to the metal surface, potentially allowing for a fine control of the relative orientation between the EEF and adsorbed molecules. 
Therefore, understanding the effects of EEFs on heterogeneous catalysis by metals is a major focus of electrostatic catalysis~\cite{Che17p551,Che17p6957,Gray20p125640}.

Methanol, as an important hydrogen-storage liquid fuel \cite{Shih18p1925}, can be used for on-board hydrogen production through steam reforming of methanol (SRM)~\cite{Trimm01p31}, in which methanol reacts with steam and produces hydrogen, CO, and CO$_2$ at about 300 $^\circ$C. However, the extensively used catalysts in industry often suffer from catalyst deactivation due to coke deposition and sintering~\cite{Sa10p43}.
Since the elementary reactions of SRM involve adsorbates possessing large electric dipole moments, we expect that an EEF will bring appreciable impact on the kinetics and thermodynamics of the SRM process as the energetics of molecules of large dipole moments are susceptible to EEFs~\cite{Wang18p13350,Shaik20p12551}. Considering the critical role of SRM played in the green methanol economy~\cite{Shih18p1925, olah05p2636}, it is imperative to design next-generation catalysts superior to current ones, particularly for low-temperature SRM. In this work, we propose to use EEF as a ``smart agent" to improve both the catalytic activity and coke resistance. 

First-principles-based computational catalyst design is emerging as a promising approach to obtain high-quality catalysts~\cite{Ke21p10860, han21p1833, fu2020p156001}. Pertinent to electrostatic catalysis, finite-field density functional theory (DFT) calculations have been used to quantify the activation energies ($\Delta E_a$) of elementary reactions on metal surfaces in the presence of EEFs of varying magnitudes, revealing important atomistic insights that helped the understanding of EEF effects~\cite{Che18p5153,Che17p6957,Chen16p7133}. However, most previous studies only focused on a few representative electric fields (\eg, negative versus positive) and their impact on the catalytic activity of a specific metal (\eg, Ni, Ag)~\cite{Che17p6957, Chen16p7133}, 
partly due to the expensive computational cost of finite-field DFT calculations. For a complex process like SRM comprising a network of elementary reactions, the EEF dependence of the overall catalytic performance is likely nonlinear such that the search of an optimal EEF should be performed in order to maximize the efficiency but has never been done. Moreover, the design of high-quality electrostatic catalysts targeting specific operation conditions requires detailed understandings of the collective effects of EEFs and various reaction parameters such as temperature, gas composition, and partial pressure. All these difficulties essentially forbid a DFT-based high-throughput computational catalyst screening in the presence of EEFs. 

\begin{figure}[htb]
    \centering
    \includegraphics[scale=0.22]{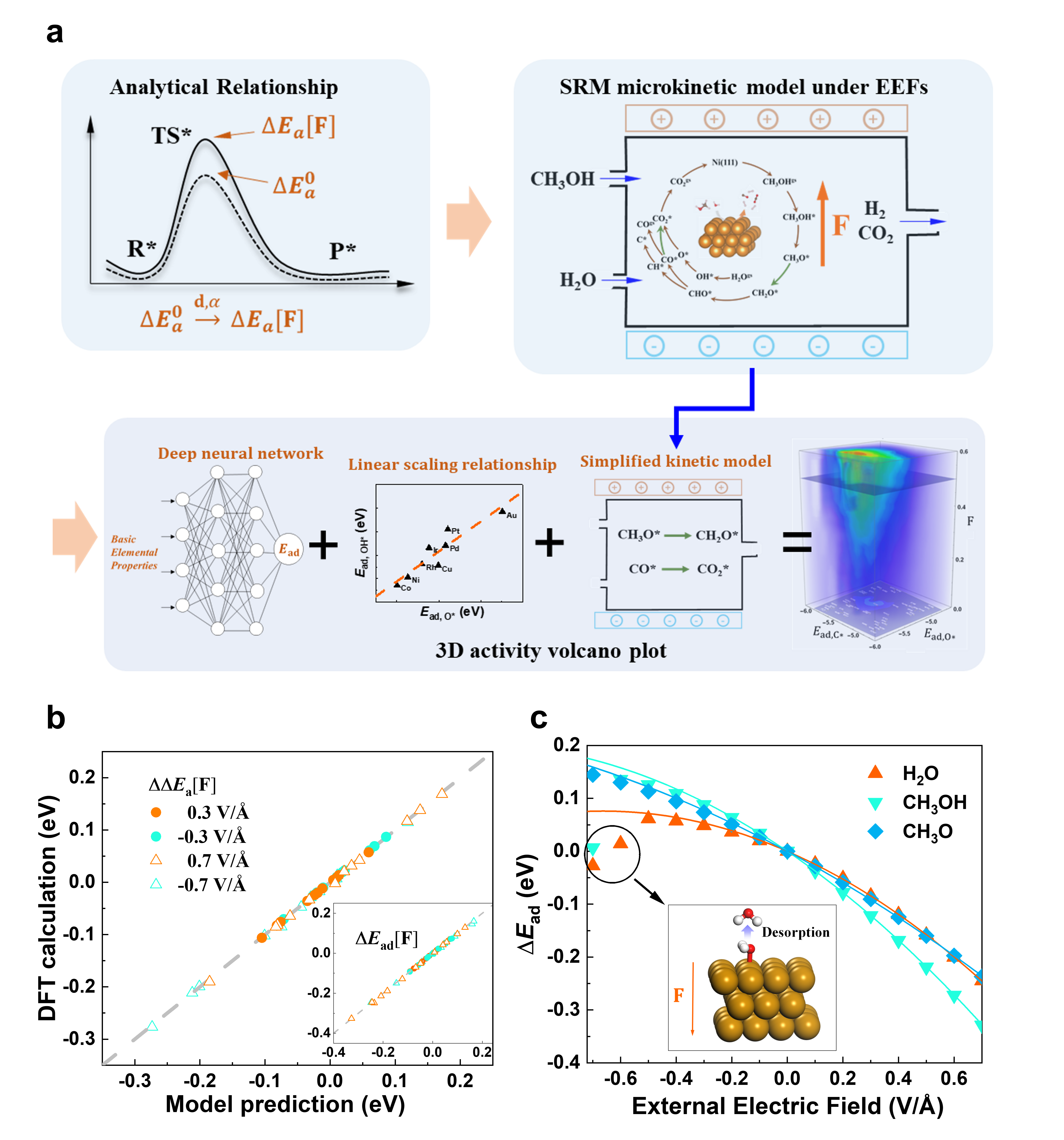}
    \caption{(a) Workflow of multiscale catalyst design under EEFs.  (b) Comparison of EEF-induced activation energy change ($\Delta \Delta E_a[\mathbf{F}]$) and adsorption energy change ($\Delta E_{\rm ad}[\mathbf{F}]$) estimated with the analytical relationships and direct DFT calculations for tightly bonded adsorbates and transition states. (c) Comparison of analytical and DFT values of $\Delta E_{\rm ad}[\mathbf{F}]$ for three representative species. The inset shows H$_2$O will desorbe from the surface due to a strong negative field.}
    \label{Fig1}
\end{figure}

Here, using SRM as an example, we develop a deep-learning-assisted first-principles-based multiscale method that leads to the {\em first} three-dimensional (3D) activity volcano plot under EEFs, quantifying the field-dependent SRM activity for thousands of metallic alloys. The multsicale framework consists of three steps, as shown in Fig.~\ref{Fig1}a. First, we derive a general analytical relationship between the electric field ($\mathbf{F}$) and $\Delta E_a$ of a surface reaction within the harmonic approximation. This enables a rapid and accurate predication of $\Delta E_a[\mathbf{F}]$ at any given field strength. The analytical relationship is then incorporated into a microkinetic mode of SRM on Ni(111) surface, resulting in a continuous stirred-tank reactor model to simulate SRM under EEFs. The microkinetic modeling reveals a highly nonlinear temperature-dependent EEF effect: a positive EEF can increase the conversion of CH$_3$OH at high temperatures ($>$ 350$^\circ$C) but suppress the conversion at low temperatures ($<$ 250$^\circ$C), highlighting the necessity of multiscale modeling for catalyst design under EEFs. Finally, by combining the linear scaling relationship~\cite{Michaelides03p3704, abild-pedersen07p016105}, a DFT-based deep neural network (DNN) that rapidly predicts C* and O* adsorption energies~\cite{Ke21p10860}, and a simplified kinetic model derived from the microkinetic model, we are able to construct the 3D activity volcano plot to identify high-quality catalysts for low-temperature SRM under EEFs.
 
{\em Method}{\indent}DFT calculations are performed using Vienna {\em ab initio} Simulation Package (VASP)~\cite{Kresse96p11169,Kresse96p15}. The interaction between core ion and electrons is described by the projector augmented wave (PAW) method~\cite{Blochl94p17953}. The Perdew-Wang-91 (PW91) functional is chosen as the exchange-correlation functional~\cite{Perdew92p13244}. The plane-wave kinetic energy cutoff is 400~eV and a 5$\times$5$\times$1 Monkhorst-Pack grid is used to sample the Brillouin zone. A three-layer 3$\times$3 Ni slab with a vacuum region of more than 10~\AA~thick is constructed to simulate the Ni(111) surface. The method proposed by Neugebauer and Scheffler~\cite{Neugebauer92p16067} is employed to apply an EEF normal to the metal surface. Climbing image nudged elastic band (CL-NEB) method \cite{Henkelman00p9901} and dimer method \cite{Henkelman99p7010} are used to identify transition states of surface reactions, and the electronic energy and atomic force are converged respectively to 10$^{-7}$ eV and 0.01 eV/\AA. More computational details can be found in Supplementary Materials.

{\em Activation and adsorption energy under EEFs within the harmonic approximation}{\indent}For an elementary surface reaction step, the EEF-induced changes in adsorption and activation energy have been attributed to various factors such as modified charge transfer between adsorbates and metal surfaces~\cite{Yeh19p8230}, bond elongation/contraction of adsorbates as well as the shift of metal work function~\cite{Pacchioni97p263}. These important yet intimately coupled atomistic mechanisms seem to suggest a quantitative determination of the EEF-induced energy change can only be achieved via genuine {\em ab initio} modeling that captures all the factors. We now show that $\Delta E_a[\mathbf{F}]$ at a given field ($\mathbf{F}$) can be readily obtained using only zero-field parameters within the harmonic approximation. 

Without loss of generality, a simple elementary reaction, R*$\rightarrow$TS*$\rightarrow$P* is considered, where R*, TS*, and P* are reactant, transition state, and product adsorbed on the metal surface, respectively, and the EEF is applied along the surface normal.  For an isolated species $i$ under an EEF, its free energy in the form of Taylor series up to the second order is 
\begin{equation}
    E_i[\mathbf{F}] = E^0 - \mathbf{d}_i\cdot \mathbf{F} -  \frac{1}{2}  \alpha_i \mathbf{F}^2,
    \label{eq1}
\end{equation}
where $E^0$ is the zero-field energy, $\mathbf{d}$ is the dipole moment, and $\alpha$ is the polarizability. Equation~\ref{eq1} applies to both isolated molecules as well as the slab model with adsorbates (denoted as $s$+$i$). It is straightforward to derive
\begin{equation}
    \Delta E_a[\mathbf{F}] = \Delta E_a^0- (\mathbf{d}_{s+\rm TS^*} - \mathbf{d}_{s+\rm R^*})\cdot  \mathbf{F} -
    \frac{1}{2} (\alpha_{s+\rm TS^*} - \alpha_{s+\rm R^*})\mathbf{F}^2
    \label{Ea}
\end{equation}
where $\Delta E^0_a = E^0_{s+\rm TS^*} - E^0_{s+\rm R^*}$ is the zero-field activation energy. We note that $\mathbf{d}_{s+\rm TS^*}$ and $\alpha_{s+\rm TS^*}$ ($\mathbf{d}_{s+\rm R^*}$ and $\alpha_{s+\rm R^*}$) are the electric dipole moment and polarizability, respectively, of the whole bounded system comprised of the adsorbed TS* (R*) and the slab used to model the surface.
It is evident that all reaction-specific parameters ($\Delta E_a^0$, $\mathbf{d}$, $\alpha$ ) in Eq.~\ref{Ea} can be computed from DFT and perturbation theory at zero field. A generalization of Eq.~\ref{Ea} gives the adsorption energy under EEFs, 
\begin{equation}
     E_{\rm ad}[\mathbf{F}] = E_{\rm ad}^0- (\mathbf{d}_{s+i} - \mathbf{d}_{s})\cdot  \mathbf{F} - \frac{1}{2} (\alpha_{s+i} - \alpha_{s})\mathbf{F}^2
    \label{ads}
\end{equation}
where $E_{\rm ad}^0 = E^0_{s+i} - E^0_{s}-E^0_{i}$. We note that our definition of adsorption energy captures the effect of EEF-driven gas diffusion  and is subtlety different from the conventional definition (see details in Supplementary Materials, Sect.~II).

Expressions similar to Eqs.~\ref{Ea}--\ref{ads} have been derived previously\cite{Che18p5153, Chen16p7133} but, surprisingly, have never been used to calculate $\Delta E_a[\mathbf{F}]$ or $E_{\rm ad}[\mathbf{F}]$, likely because of the general assumption that EEF-induced structural and electronic changes of metal surface-adsorbate complex would be highly anharmonic thus beyond the applicability of harmonic approximation. We compare the EEF-induced energy change, $\Delta \Delta E_a[\mathbf{F}] = \Delta E_a[\mathbf{F}] - \Delta E_a^0 $ and $\Delta  E_{\rm ad}[\mathbf{F}] = E_{\rm ad}[\mathbf{F}] - E_{\rm ad}^0$, obtained with Eqs.~\ref{Ea}--\ref{ads} using only zero-field parameters (see values in Supplementary Materials, Sect.~III) and those from direct DFT calculations in Fig.\ref{Fig1}b. With a mean absolute error of only 1 meV and 2 meV for $\Delta \Delta E_a[\mathbf{F}]$ and $\Delta  E_{\rm ad}[\mathbf{F}]$, respectively, the simple analytical relationships described by Eqs.~\ref{Ea}--\ref{ads} have remarkable accuracy over a wide range of field strengths. 
We further compare the analytical and DFT values of $\Delta  E_{\rm ad}[\mathbf{F}]$ for three representative molecules, H$_2$O, CH$_3$OH, and CH$_3$O in Fig.\ref{Fig1}c, and find a
satisfying agreement for $-0.4<\mathbf{F}<0.7$~eV/\AA.
Consistent with previous DFT investigations~\cite{Che18p5153, Che16p77}, a negative field will induce desorption ($\Delta  E_{\rm ad}[\mathbf{F}]>0$) while a positive field may facilitate the adsorption ($\Delta  E_{\rm ad}[\mathbf{F}]<0$) for a molecule with negative electric dipole moment on the metal surface. It is only when the molecules become desorbed due to a strong negative field (\eg, $-0.5$~V/\AA~for H$_2$O), the analytical value of $\Delta  E_{\rm ad}[\mathbf{F}]$ starts deviating from the DFT result. 

The demonstrated quantitative nature of a rather simple analytical theory linking field strength to EEF-induced energy change seems counterintuitive. From Eqs.~\ref{Ea}--\ref{ads}, it is clear that only the whole bounded system (the slab with an adsorbed molecule) is relevant to EEF-dependent terms. 
We argue that for the the whole bounded system, the impact of an EEF on the total energy is a small perturbation that could be well described within the harmonic approximation, despite highly anharmonic local changes between the adsorbate and the metal surface (see numerical verification in Fig.~S3). 
\begin{figure}[b]
    \centering
    \includegraphics[scale=0.22]{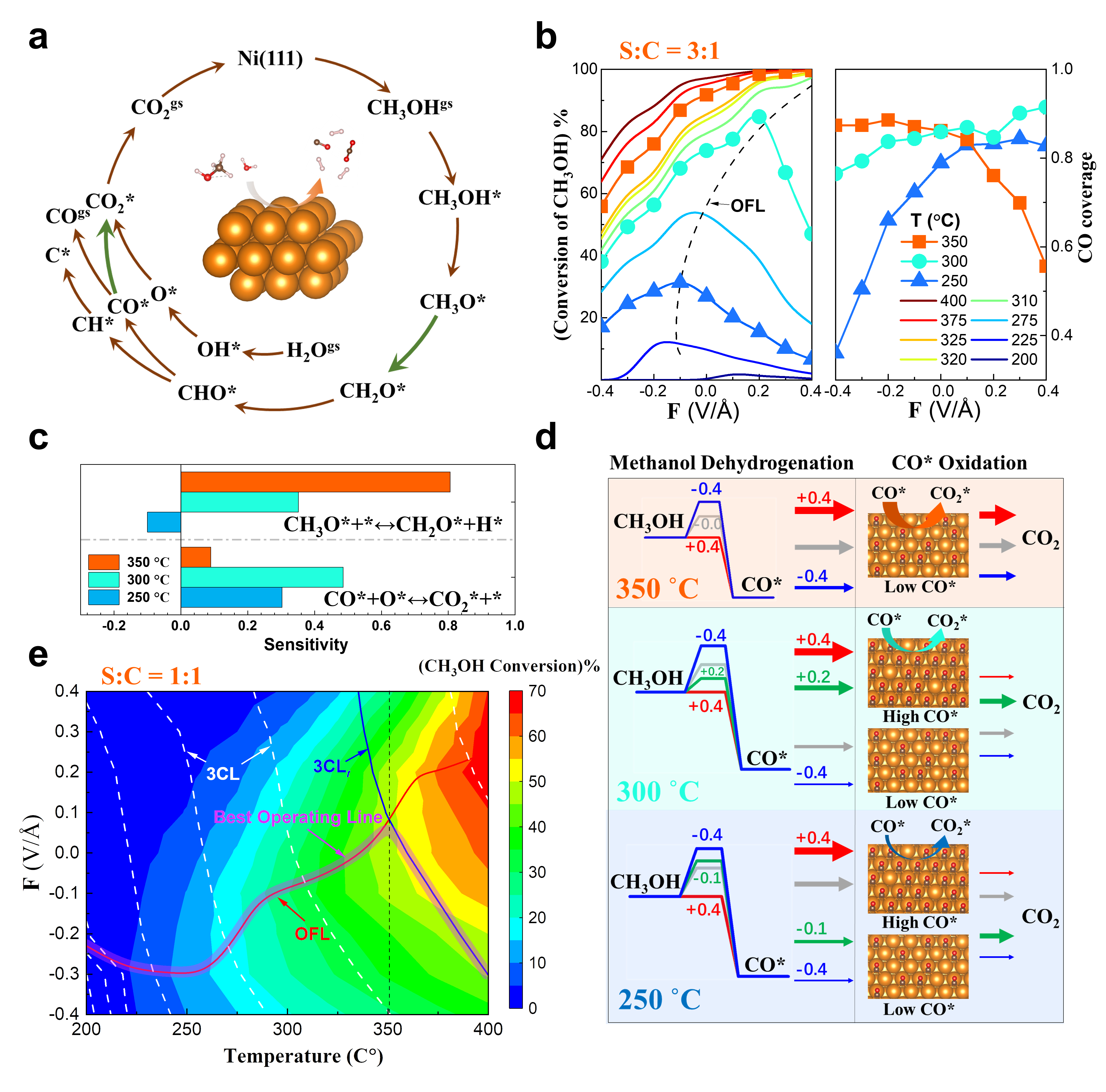}
    \caption{Nonlinear temperature-dependent EEF effect in SRM. (a) Reaction network of SRM considered in this work. The green arrows highlight two RDSs. (b) Methanol conversion rate (left) and CO* coverage (right) as a function of electric field strength at different temperatures for S:M = 3:1. (c) Sensitivity analysis of SRM. (d) Impacts of EEFs on the two RDSs at 350$^\circ$C, 300$^\circ$C, and 250$^\circ$C. The width of the arrow scales with the rate. (e) Countour plot of methanol conversion rate as a function of temperature and field strength for S:M = 1:1.}
    \label{Fig2}
\end{figure}

{\em Microkinetics of SRM under EEFs}{\indent}The reaction network of SRM, illustrated in Fig.\ref{Fig2}a, includes CH$_3$OH and H$_2$O dehydrogenation and CO* oxidation. Based on Eqs.~\ref{Ea}--\ref{ads}, we construct a microkinetic model that simulates the Ni-catalyzed SRM process in the presence of EEFs under realistic conditions that take into account the impacts of gas composition, temperature, and inlet velocity. The validity of the microkinetic model is confirmed by comparing theoretical predictions with available experimental results (see details in Supplementary Materials, Sect. IV). For a 3:1 steam to methanol ratio (S:M = 3:1), the temperature- and field-dependence of the conversion of CH$_3$OH is presented in Fig.~\ref{Fig2}b where we define an ``optimal field line" (OFL) that tracks the electric field resulting in the highest conversion. The OFL reveals several interesting characteristics of SRM under EEFs. 
At high temperatures $\ge 310^{\circ}$C, a positive field can promote the overall reaction rate compared to the zero-field case, and the degree of enhancement scales with the field strength. Interestingly, at an intermediate temperature such as $300^{\circ}$C, a large positive field that is to the right of the OFL will suppress the conversion. When the OFL crosses the zero field and enters into the negative field region, it means a negative field now can facilitate the conversion; this happens at $\approx$250~$^{\circ}$C, but a too strong negative field also becomes unfavorable.

We perform sensitivity analysis by computing the partial derivatives of methanol conversion rate with respect to the rate constant of the elementary reaction~\cite{Campbell21p647}. As shown in Fig.~\ref{Fig2}c, there are two RDSs in SRM, the dehydrogenation of CH$_3$O*, CH$_3$O*+* $\rightarrow$ CH$_2$O*+H*, and the consumption of CO*, CO*+O*$\rightarrow$CO$_2$. The nontrivial evolution of OFL is due to a delicate balance between the two RDSs with the dehydrogenation of CH$_3$O* being more EEF sensitive while the consumption of CO* being more temperature sensitive.  Specifically, as shown in Fig.~\ref{Fig2}d, at a high temperature of $350^{\circ}$C, the overall reaction rate of SRM is dictated by the dehydrogenation of CH$_3$O* as the consumption of CO* is fast enough at such high temperature. Therefore, a stronger positive EEF that reduces the activation energy of CH$_3$O* dehydration leads to increased methanol conversion.
When the temperature is reduced to $300^{\circ}$C, the consumption of CO* remains fast such that applying a positive EEF below $\mathbf{F}_{\rm OFL}$ can still facilitate the conversion by increasing the rate of CH$_3$O* dehydrogenation. However, after passing $\mathbf{F}_{\rm OFL}$, the CO* generation will outspeed the consumption, causing CO poisoning of Ni. At a low temperature of $250^{\circ}$C, the CO* consumption becomes rather slow thus requiring a negative field to suppress CH$_3$O* dehydration to inhibit CO poisoning; a too strong negative field that severely slows down CH$_3$O* dehydration will unsurprisingly cause a low methanol conversion rate. Simply put, the OFL essentially marks the boundary between CH$_3$OH-dehydrogenation-controlled region and CO-poisoning-controlled region.

It is well known that coking is the most common deactivation mechanism of industrial SRM catalysts~\cite{Valdes06p354,Twigg03p191}, and increasing the steam pressure is the general approach to suppress coking. Though a high S:M ratio (\eg, 3:1) is beneficial for enhancing coke resistance, it will raise energy consumption for steam heating, reduce the methanol conversion rate, and decrease the H$_2$ content of final products. It would be helpful to have an efficient tool to identify the optimal conditions for balanced SRM rate and coke resistance. 
For S:M = 1:1, we construct a contour plot of methanol conversion rate as a function of temperature and EEF strength (Fig.~\ref{Fig2}e), showing both OFL and constant carbon concentration lines (3CLs). For a target level of coke resistance (a selected 3CL, denoted as 3CL$_t$), the lower part of the merged OFL and 3CL$_t$ gives the best operating EEF for a wide range of temperatures.

{\em 3D activity volcano plot under EEFs}{\indent}A detailed study of the microkinetic model of SRM shows that the methanol conversion rate ($r_{\rm M}[\mathbf{F}]$) can be quantified with a simplified pathway~\cite{Ke20p349},
\[ {\rm CH_3OH} \leftrightarrows {\rm CH_3OH*} \leftrightarrows {\rm CH_3O*} \stackrel{k^{\rm RDS}}{\longrightarrow} {\rm CH_2O*} \]
in which CH$_3$O*+* $\rightarrow$ CH$_2$O*+H* is the RDS. The corresponding rate equation reads
\begin{equation}
{r_{\rm M}[\mathbf{F}] =  A \cdot \exp(-\frac{\Delta E_a^{\rm RDS}[\mathbf{F}]}{RT}) \frac{P_{\rm CH_3OH}[\mathbf{F}]}{\sqrt{P_{\rm H_2}[\mathbf{F}]}} \cdot \theta[\mathbf{F}]^2.}
\label{rate}
\end{equation}
Here, $P_{\rm CH_3OH}[\mathbf{F}]$, $P_{\rm H_2}[\mathbf{F}]$, $\theta[\mathbf{F}]$ are the field-dependent partial pressure of CH$_3$OH, H$_2$, and the coverage of active sites that depend on the adsorption energies of CO*, O*, and H* (see Section VI in Supplementary Materials); $A$ is a temperature dependent parameter deduced from DFT calculations. The simplified kinetic model based on Eq.~\ref{rate} is capable of reproducing the results of the microkinetic model based on the complete reaction network but with a fraction of the computational cost.

\begin{figure}
    \centering
    \includegraphics[scale=0.22]{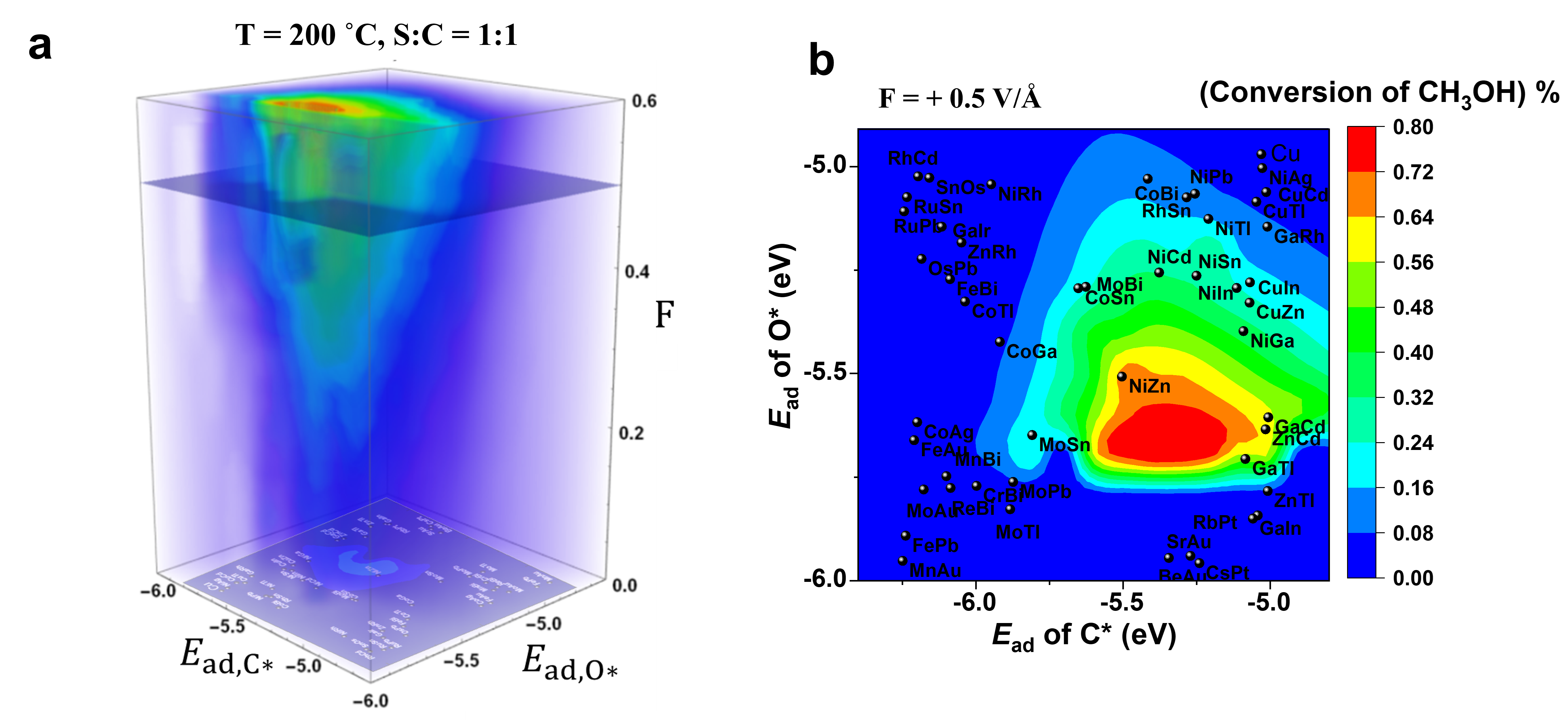}
    \caption{ (a) 3D activity volcano plot for SRM at 200~$^\circ$C and S:M = 1:1 under EEFs with a slice shown in (b).}
    \label{3D}
\end{figure}

To construct the activity volcano plot under EEFs for a large number of metallic catalysts, accurate and rapid determinations of $E_{\rm ad}[\mathbf{F}]$ and $\Delta E_a[\mathbf{F}]$ are prerequisite. We find that the EEF-induced energy change turns out to be rather insensitive to the metal type, which greatly simplifies the problem (see Supplementary Material Sect. VII). That is, for an adsorbate $i$ on a given metal surface ($m$), the field-dependent adsorption energy is reduced to $E_{{\rm ad}}[\mathbf{F},m] = E_{{\rm ad}}^0[m] -( \mathbf{d}_{s_0+i}- \mathbf{d}_{s_0})\cdot  \mathbf{F} - \frac{1}{2} (\alpha_{s_0+i} - \alpha_{s_0})\mathbf{F}^2$, where $E_{{\rm ad}}^0[m]$ is the zero-field adsorption energy on metal $m$ that can be estimated using C* and O* adsorption energies on metal $m$ with the well-known linear scaling relationship~\cite{Ke21p10860} while the last two terms only depend on $\mathbf{F}$ and are computed using a reference metal surface $s_0$. 
Equipped with a DNN that can quickly predict the zero-field adsorption energies of C* and O*~\cite{Ke21p10860} for various metallic alloys and the simplified kinetic model (Eq.~\ref{rate}), we finally construct a 3D activity volcano plot that quantitatively predicts the menthol conversion in the presence of EEFs for 1711 metallic alloys (Fig.~\ref{3D}a) at a low temperature of 200~$^\circ$C. This 3D volcano plot contains rich information and is worthy of detailed investigations. Here we only highlight the catalytic performance of NiZn. Despite having a low SRM activity at zero field, NiZn can promote the methanol conversion in the presence of an EEF of 0.5~V/\AA~at 200~$^\circ$C. Combined with its proved coke resistance~\cite{Ke21p10860}, EEF-assisted NiZn is recommended as an efficient and  environment-friendly catalyst for SRM at 200~$^\circ$C and S:M = 1:1. 

%\section{Conclusions}
{\em Conclusion.}{\indent}In summary, we demonstrate that a simple analytical theory within the harmonic approximation can accurately predict finite-field energetics using only zero-field parameters, eliminating the needs of expensive finite-field DFT calculations for field-dependent microkinetic modeling. 
Focusing on an important process in green methanol economy, steam reforming of methanol, we reveal nontrivial collective effects of EEF and temperature: a positive EEF enhances the methanol conversion at higher temperatures while suppressing the overall reaction at lower temperatures. The introduction of OFL and 3CL$_t$ in the temperature--EEF parameter space allows for facile determinations of best operating EEF for a wide range of temperatures. Finally, using a chemistry-based simplified kinetic model and a first-principles-based DNN, we construct a 3D activity volcano plot under EEFs for 1711 metallic alloys and identify EEF-assisted NiZn for low-temperature steam reforming of methanol. We expect the multiscale approach developed in this work can be readily applied to other heterogeneous catalysis by metal under EEFs. 

\begin{acknowledgments}
C.K., S.L. acknowledge the supports from Westlake Education Foundation. C. K. acknowledge the help from Yudi Yang during the preparation of the manuscript. The computational resource is provided by Westlake HPC Center.
\end{acknowledgments}
%{\bf{Competing Interests}} The authors declare no competing financial or non-financial interests.

%{\bf{Data Availability}} The data that support the findings of this study are included in this article and are available from the corresponding author upon reasonable request.

%{\bf{Author Contributions}} S.L. proposed the design principle and led the project. J.H., X.D., S.J., Y.K., and S.L. carried out first-principles DFT calculations. J.L. developed the analytical model. J.Z. proposed the bias-free optospintronics. All authors contributed to the discussion of the manuscript. 

\newpage
\bibliography{SL}% Produces the bibliography via BibTeX.

\begin{thebibliography}{38}%
\makeatletter
\providecommand \@ifxundefined [1]{%
 \@ifx{#1\undefined}
}%
\providecommand \@ifnum [1]{%
 \ifnum #1\expandafter \@firstoftwo
 \else \expandafter \@secondoftwo
 \fi
}%
\providecommand \@ifx [1]{%
 \ifx #1\expandafter \@firstoftwo
 \else \expandafter \@secondoftwo
 \fi
}%
\providecommand \natexlab [1]{#1}%
\providecommand \enquote  [1]{``#1''}%
\providecommand \bibnamefont  [1]{#1}%
\providecommand \bibfnamefont [1]{#1}%
\providecommand \citenamefont [1]{#1}%
\providecommand \href@noop [0]{\@secondoftwo}%
\providecommand \href [0]{\begingroup \@sanitize@url \@href}%
\providecommand \@href[1]{\@@startlink{#1}\@@href}%
\providecommand \@@href[1]{\endgroup#1\@@endlink}%
\providecommand \@sanitize@url [0]{\catcode `\\12\catcode `\$12\catcode
  `\&12\catcode `\#12\catcode `\^12\catcode `\_12\catcode `\%12\relax}%
\providecommand \@@startlink[1]{}%
\providecommand \@@endlink[0]{}%
\providecommand \url  [0]{\begingroup\@sanitize@url \@url }%
\providecommand \@url [1]{\endgroup\@href {#1}{\urlprefix }}%
\providecommand \urlprefix  [0]{URL }%
\providecommand \Eprint [0]{\href }%
\providecommand \doibase [0]{https://doi.org/}%
\providecommand \selectlanguage [0]{\@gobble}%
\providecommand \bibinfo  [0]{\@secondoftwo}%
\providecommand \bibfield  [0]{\@secondoftwo}%
\providecommand \translation [1]{[#1]}%
\providecommand \BibitemOpen [0]{}%
\providecommand \bibitemStop [0]{}%
\providecommand \bibitemNoStop [0]{.\EOS\space}%
\providecommand \EOS [0]{\spacefactor3000\relax}%
\providecommand \BibitemShut  [1]{\csname bibitem#1\endcsname}%
\let\auto@bib@innerbib\@empty
%</preamble>
\bibitem [{\citenamefont {Shaik}\ \emph {et~al.}(2016)\citenamefont {Shaik},
  \citenamefont {Mandal},\ and\ \citenamefont {Ramanan}}]{Shaik16p1091}%
  \BibitemOpen
  \bibfield  {author} {\bibinfo {author} {\bibfnamefont {S.}~\bibnamefont
  {Shaik}}, \bibinfo {author} {\bibfnamefont {D.}~\bibnamefont {Mandal}},\ and\
  \bibinfo {author} {\bibfnamefont {R.}~\bibnamefont {Ramanan}},\ }\bibfield
  {title} {\bibinfo {title} {Oriented electric fields as future smart reagents
  in chemistry},\ }\href@noop {} {\bibfield  {journal} {\bibinfo  {journal}
  {Nat. Chem.}\ }\textbf {\bibinfo {volume} {8}},\ \bibinfo {pages} {1091}
  (\bibinfo {year} {2016})}\BibitemShut {NoStop}%
\bibitem [{\citenamefont {Shaik}\ \emph {et~al.}(2018)\citenamefont {Shaik},
  \citenamefont {Ramanan}, \citenamefont {Danovich},\ and\ \citenamefont
  {Mandal}}]{Shaik18p5125}%
  \BibitemOpen
  \bibfield  {author} {\bibinfo {author} {\bibfnamefont {S.}~\bibnamefont
  {Shaik}}, \bibinfo {author} {\bibfnamefont {R.}~\bibnamefont {Ramanan}},
  \bibinfo {author} {\bibfnamefont {D.}~\bibnamefont {Danovich}},\ and\
  \bibinfo {author} {\bibfnamefont {D.}~\bibnamefont {Mandal}},\ }\bibfield
  {title} {\bibinfo {title} {Structure and reactivity/selectivity control by
  oriented-external electric fields},\ }\href
  {https://doi.org/10.1039/C8CS00354H} {\bibfield  {journal} {\bibinfo
  {journal} {Chem. Soc. Rev.}\ }\textbf {\bibinfo {volume} {47}},\ \bibinfo
  {pages} {5125} (\bibinfo {year} {2018})}\BibitemShut {NoStop}%
\bibitem [{\citenamefont {Che}\ \emph {et~al.}(2018)\citenamefont {Che},
  \citenamefont {Gray}, \citenamefont {Ha}, \citenamefont {Kruse},
  \citenamefont {Scott},\ and\ \citenamefont {{McEwen}}}]{Che18p5153}%
  \BibitemOpen
  \bibfield  {author} {\bibinfo {author} {\bibfnamefont {F.}~\bibnamefont
  {Che}}, \bibinfo {author} {\bibfnamefont {J.~T.}\ \bibnamefont {Gray}},
  \bibinfo {author} {\bibfnamefont {S.}~\bibnamefont {Ha}}, \bibinfo {author}
  {\bibfnamefont {N.}~\bibnamefont {Kruse}}, \bibinfo {author} {\bibfnamefont
  {S.~L.}\ \bibnamefont {Scott}},\ and\ \bibinfo {author} {\bibfnamefont
  {J.-S.}\ \bibnamefont {{McEwen}}},\ }\bibfield  {title} {\bibinfo {title}
  {Elucidating the roles of electric fields in catalysis\: A perspective},\
  }\href@noop {} {\bibfield  {journal} {\bibinfo  {journal} {ACS Catal.}\
  }\textbf {\bibinfo {volume} {8}},\ \bibinfo {pages} {5153} (\bibinfo {year}
  {2018})}\BibitemShut {NoStop}%
\bibitem [{\citenamefont {Ciampi}\ \emph {et~al.}(2018)\citenamefont {Ciampi},
  \citenamefont {Darwish}, \citenamefont {Aitken}, \citenamefont
  {Díez-Pérez},\ and\ \citenamefont {Coote}}]{Ciampi18p5146}%
  \BibitemOpen
  \bibfield  {author} {\bibinfo {author} {\bibfnamefont {S.}~\bibnamefont
  {Ciampi}}, \bibinfo {author} {\bibfnamefont {N.}~\bibnamefont {Darwish}},
  \bibinfo {author} {\bibfnamefont {H.~M.}\ \bibnamefont {Aitken}}, \bibinfo
  {author} {\bibfnamefont {I.}~\bibnamefont {Díez-Pérez}},\ and\ \bibinfo
  {author} {\bibfnamefont {M.~L.}\ \bibnamefont {Coote}},\ }\bibfield  {title}
  {\bibinfo {title} {Harnessing electrostatic catalysis in single molecule,
  electrochemical and chemical systems: a rapidly growing experimental tool
  box},\ }\href@noop {} {\bibfield  {journal} {\bibinfo  {journal} {Chem. Soc.
  Rev.}\ }\textbf {\bibinfo {volume} {47}},\ \bibinfo {pages} {5146} (\bibinfo
  {year} {2018})}\BibitemShut {NoStop}%
\bibitem [{\citenamefont {Shaik}\ \emph {et~al.}(2020)\citenamefont {Shaik},
  \citenamefont {Danovich}, \citenamefont {Joy}, \citenamefont {Wang},\ and\
  \citenamefont {Stuyver}}]{Shaik20p12551}%
  \BibitemOpen
  \bibfield  {author} {\bibinfo {author} {\bibfnamefont {S.}~\bibnamefont
  {Shaik}}, \bibinfo {author} {\bibfnamefont {D.}~\bibnamefont {Danovich}},
  \bibinfo {author} {\bibfnamefont {J.}~\bibnamefont {Joy}}, \bibinfo {author}
  {\bibfnamefont {Z.}~\bibnamefont {Wang}},\ and\ \bibinfo {author}
  {\bibfnamefont {T.}~\bibnamefont {Stuyver}},\ }\bibfield  {title} {\bibinfo
  {title} {Electric-field mediated chemistry: Uncovering and exploiting the
  potential of (oriented) electric fields to exert chemical catalysis and
  reaction control},\ }\href@noop {} {\bibfield  {journal} {\bibinfo  {journal}
  {J. Am. Chem. Soc.}\ }\textbf {\bibinfo {volume} {142}},\ \bibinfo {pages}
  {12551} (\bibinfo {year} {2020})}\BibitemShut {NoStop}%
\bibitem [{\citenamefont {Stuyver}\ \emph {et~al.}(2020)\citenamefont
  {Stuyver}, \citenamefont {Danovich}, \citenamefont {Joy},\ and\ \citenamefont
  {Shaik}}]{Stuyver20pe1438}%
  \BibitemOpen
  \bibfield  {author} {\bibinfo {author} {\bibfnamefont {T.}~\bibnamefont
  {Stuyver}}, \bibinfo {author} {\bibfnamefont {D.}~\bibnamefont {Danovich}},
  \bibinfo {author} {\bibfnamefont {J.}~\bibnamefont {Joy}},\ and\ \bibinfo
  {author} {\bibfnamefont {S.}~\bibnamefont {Shaik}},\ }\bibfield  {title}
  {\bibinfo {title} {External electric field effects on chemical structure and
  reactivity},\ }\href@noop {} {\bibfield  {journal} {\bibinfo  {journal}
  {{WIREs} Comput. Mol. Sci.}\ }\textbf {\bibinfo {volume} {10}},\ \bibinfo
  {pages} {e1438} (\bibinfo {year} {2020})}\BibitemShut {NoStop}%
\bibitem [{\citenamefont {Aragonès}\ \emph {et~al.}(2016)\citenamefont
  {Aragonès}, \citenamefont {Haworth}, \citenamefont {Darwish}, \citenamefont
  {Ciampi}, \citenamefont {Bloomfield}, \citenamefont {Wallace}, \citenamefont
  {Diez-Perez},\ and\ \citenamefont {Coote}}]{Aragones16p88}%
  \BibitemOpen
  \bibfield  {author} {\bibinfo {author} {\bibfnamefont {A.~C.}\ \bibnamefont
  {Aragonès}}, \bibinfo {author} {\bibfnamefont {N.~L.}\ \bibnamefont
  {Haworth}}, \bibinfo {author} {\bibfnamefont {N.}~\bibnamefont {Darwish}},
  \bibinfo {author} {\bibfnamefont {S.}~\bibnamefont {Ciampi}}, \bibinfo
  {author} {\bibfnamefont {N.~J.}\ \bibnamefont {Bloomfield}}, \bibinfo
  {author} {\bibfnamefont {G.~G.}\ \bibnamefont {Wallace}}, \bibinfo {author}
  {\bibfnamefont {I.}~\bibnamefont {Diez-Perez}},\ and\ \bibinfo {author}
  {\bibfnamefont {M.~L.}\ \bibnamefont {Coote}},\ }\bibfield  {title} {\bibinfo
  {title} {Electrostatic catalysis of a diels–alder reaction},\ }\href
  {https://doi.org/10.1038/nature16989} {\bibfield  {journal} {\bibinfo
  {journal} {Nature}\ }\textbf {\bibinfo {volume} {531}},\ \bibinfo {pages}
  {88} (\bibinfo {year} {2016})}\BibitemShut {NoStop}%
\bibitem [{\citenamefont {Oshima}\ \emph {et~al.}(2013)\citenamefont {Oshima},
  \citenamefont {Shinagawa}, \citenamefont {Haraguchi},\ and\ \citenamefont
  {Sekine}}]{Oshima13p3003}%
  \BibitemOpen
  \bibfield  {author} {\bibinfo {author} {\bibfnamefont {K.}~\bibnamefont
  {Oshima}}, \bibinfo {author} {\bibfnamefont {T.}~\bibnamefont {Shinagawa}},
  \bibinfo {author} {\bibfnamefont {M.}~\bibnamefont {Haraguchi}},\ and\
  \bibinfo {author} {\bibfnamefont {Y.}~\bibnamefont {Sekine}},\ }\bibfield
  {title} {\bibinfo {title} {Low temperature hydrogen production by catalytic
  steam reforming of methane in an electric field},\ }\href@noop {} {\bibfield
  {journal} {\bibinfo  {journal} {Int. J. Hydrogen Energ.}\ }\textbf {\bibinfo
  {volume} {38}},\ \bibinfo {pages} {3003} (\bibinfo {year}
  {2013})}\BibitemShut {NoStop}%
\bibitem [{\citenamefont {Oshima}\ \emph {et~al.}(2014)\citenamefont {Oshima},
  \citenamefont {Shinagawa}, \citenamefont {Nogami}, \citenamefont {Manabe},
  \citenamefont {Ogo},\ and\ \citenamefont {Sekine}}]{Oshima14p27}%
  \BibitemOpen
  \bibfield  {author} {\bibinfo {author} {\bibfnamefont {K.}~\bibnamefont
  {Oshima}}, \bibinfo {author} {\bibfnamefont {T.}~\bibnamefont {Shinagawa}},
  \bibinfo {author} {\bibfnamefont {Y.}~\bibnamefont {Nogami}}, \bibinfo
  {author} {\bibfnamefont {R.}~\bibnamefont {Manabe}}, \bibinfo {author}
  {\bibfnamefont {S.}~\bibnamefont {Ogo}},\ and\ \bibinfo {author}
  {\bibfnamefont {Y.}~\bibnamefont {Sekine}},\ }\bibfield  {title} {\bibinfo
  {title} {Low temperature catalytic reverse water gas shift reaction assisted
  by an electric field},\ }\href@noop {} {\bibfield  {journal} {\bibinfo
  {journal} {Catal. Today}\ }\textbf {\bibinfo {volume} {232}},\ \bibinfo
  {pages} {27} (\bibinfo {year} {2014})}\BibitemShut {NoStop}%
\bibitem [{\citenamefont {Sekine}\ \emph {et~al.}(2009)\citenamefont {Sekine},
  \citenamefont {Tomioka}, \citenamefont {Matsukata},\ and\ \citenamefont
  {Kikuchi}}]{Sekine09p183}%
  \BibitemOpen
  \bibfield  {author} {\bibinfo {author} {\bibfnamefont {Y.}~\bibnamefont
  {Sekine}}, \bibinfo {author} {\bibfnamefont {M.}~\bibnamefont {Tomioka}},
  \bibinfo {author} {\bibfnamefont {M.}~\bibnamefont {Matsukata}},\ and\
  \bibinfo {author} {\bibfnamefont {E.}~\bibnamefont {Kikuchi}},\ }\bibfield
  {title} {\bibinfo {title} {Catalytic degradation of ethanol in an electric
  field},\ }\href@noop {} {\bibfield  {journal} {\bibinfo  {journal} {Catal.
  Today}\ }\textbf {\bibinfo {volume} {146}},\ \bibinfo {pages} {183} (\bibinfo
  {year} {2009})}\BibitemShut {NoStop}%
\bibitem [{\citenamefont {Gray}\ \emph {et~al.}(2020)\citenamefont {Gray},
  \citenamefont {Agarwal}, \citenamefont {Cho}, \citenamefont {Yang},\ and\
  \citenamefont {Ha}}]{Gray20p125640}%
  \BibitemOpen
  \bibfield  {author} {\bibinfo {author} {\bibfnamefont {J.~T.}\ \bibnamefont
  {Gray}}, \bibinfo {author} {\bibfnamefont {K.}~\bibnamefont {Agarwal}},
  \bibinfo {author} {\bibfnamefont {J.-H.}\ \bibnamefont {Cho}}, \bibinfo
  {author} {\bibfnamefont {J.-I.}\ \bibnamefont {Yang}},\ and\ \bibinfo
  {author} {\bibfnamefont {S.}~\bibnamefont {Ha}},\ }\bibfield  {title}
  {\bibinfo {title} {Estimating surface electric fields using reactive formic
  acid probes and {SEM} image brightness analysis},\ }\href@noop {} {\bibfield
  {journal} {\bibinfo  {journal} {Chem. Eng. J.}\ }\textbf {\bibinfo {volume}
  {402}},\ \bibinfo {pages} {125640} (\bibinfo {year} {2020})}\BibitemShut
  {NoStop}%
\bibitem [{\citenamefont {Che}\ \emph {et~al.}(2017{\natexlab{a}})\citenamefont
  {Che}, \citenamefont {Gray}, \citenamefont {Ha},\ and\ \citenamefont
  {{McEwen}}}]{Che17p551}%
  \BibitemOpen
  \bibfield  {author} {\bibinfo {author} {\bibfnamefont {F.}~\bibnamefont
  {Che}}, \bibinfo {author} {\bibfnamefont {J.~T.}\ \bibnamefont {Gray}},
  \bibinfo {author} {\bibfnamefont {S.}~\bibnamefont {Ha}},\ and\ \bibinfo
  {author} {\bibfnamefont {J.-S.}\ \bibnamefont {{McEwen}}},\ }\bibfield
  {title} {\bibinfo {title} {Improving ni catalysts using electric fields: A
  {DFT} and experimental study of the methane steam reforming reaction},\
  }\href {https://doi.org/10.1021/acscatal.6b02318} {\bibfield  {journal}
  {\bibinfo  {journal} {ACS Catal.}\ }\textbf {\bibinfo {volume} {7}},\
  \bibinfo {pages} {551} (\bibinfo {year} {2017}{\natexlab{a}})}\BibitemShut
  {NoStop}%
\bibitem [{\citenamefont {Che}\ \emph {et~al.}(2017{\natexlab{b}})\citenamefont
  {Che}, \citenamefont {Gray}, \citenamefont {Ha},\ and\ \citenamefont
  {{McEwen}}}]{Che17p6957}%
  \BibitemOpen
  \bibfield  {author} {\bibinfo {author} {\bibfnamefont {F.}~\bibnamefont
  {Che}}, \bibinfo {author} {\bibfnamefont {J.~T.}\ \bibnamefont {Gray}},
  \bibinfo {author} {\bibfnamefont {S.}~\bibnamefont {Ha}},\ and\ \bibinfo
  {author} {\bibfnamefont {J.-S.}\ \bibnamefont {{McEwen}}},\ }\bibfield
  {title} {\bibinfo {title} {Reducing reaction temperature, steam requirements,
  and coke formation during methane steam reforming using electric fields: A
  microkinetic modeling and experimental study},\ }\href@noop {} {\bibfield
  {journal} {\bibinfo  {journal} {{ACS} Catal.}\ }\textbf {\bibinfo {volume}
  {7}},\ \bibinfo {pages} {6957} (\bibinfo {year}
  {2017}{\natexlab{b}})}\BibitemShut {NoStop}%
\bibitem [{\citenamefont {Shih}\ \emph {et~al.}(2018)\citenamefont {Shih},
  \citenamefont {Zhang}, \citenamefont {Li},\ and\ \citenamefont
  {Bai}}]{Shih18p1925}%
  \BibitemOpen
  \bibfield  {author} {\bibinfo {author} {\bibfnamefont {C.~F.}\ \bibnamefont
  {Shih}}, \bibinfo {author} {\bibfnamefont {T.}~\bibnamefont {Zhang}},
  \bibinfo {author} {\bibfnamefont {J.}~\bibnamefont {Li}},\ and\ \bibinfo
  {author} {\bibfnamefont {C.}~\bibnamefont {Bai}},\ }\bibfield  {title}
  {\bibinfo {title} {Powering the future with liquid sunshine},\ }\href@noop {}
  {\bibfield  {journal} {\bibinfo  {journal} {Joule}\ }\textbf {\bibinfo
  {volume} {2}},\ \bibinfo {pages} {1925} (\bibinfo {year} {2018})}\BibitemShut
  {NoStop}%
\bibitem [{\citenamefont {Trimm}\ and\ \citenamefont
  {Önsan}(2001)}]{Trimm01p31}%
  \BibitemOpen
  \bibfield  {author} {\bibinfo {author} {\bibfnamefont {D.~L.}\ \bibnamefont
  {Trimm}}\ and\ \bibinfo {author} {\bibfnamefont {Z.~I.}\ \bibnamefont
  {Önsan}},\ }\bibfield  {title} {\bibinfo {title} {Onboard fuel conversion
  for hydrogen-fuel-cell-driven vehicles},\ }\href
  {https://doi.org/10.1081/CR-100104386} {\bibfield  {journal} {\bibinfo
  {journal} {Catal. Rev.}\ }\textbf {\bibinfo {volume} {43}},\ \bibinfo {pages}
  {31} (\bibinfo {year} {2001})}\BibitemShut {NoStop}%
\bibitem [{\citenamefont {Sá}\ \emph {et~al.}(2010)\citenamefont {Sá},
  \citenamefont {Silva}, \citenamefont {Brandão}, \citenamefont {Sousa},\ and\
  \citenamefont {Mendes}}]{Sa10p43}%
  \BibitemOpen
  \bibfield  {author} {\bibinfo {author} {\bibfnamefont {S.}~\bibnamefont
  {Sá}}, \bibinfo {author} {\bibfnamefont {H.}~\bibnamefont {Silva}}, \bibinfo
  {author} {\bibfnamefont {L.}~\bibnamefont {Brandão}}, \bibinfo {author}
  {\bibfnamefont {J.~M.}\ \bibnamefont {Sousa}},\ and\ \bibinfo {author}
  {\bibfnamefont {A.}~\bibnamefont {Mendes}},\ }\bibfield  {title} {\bibinfo
  {title} {Catalysts for methanol steam reforming—a review},\ }\href@noop {}
  {\bibfield  {journal} {\bibinfo  {journal} {Applied Catalysis B:
  Environmental}\ }\textbf {\bibinfo {volume} {99}},\ \bibinfo {pages} {43}
  (\bibinfo {year} {2010})}\BibitemShut {NoStop}%
\bibitem [{\citenamefont {Wang}\ \emph {et~al.}(2018)\citenamefont {Wang},
  \citenamefont {Danovich}, \citenamefont {Ramanan},\ and\ \citenamefont
  {Shaik}}]{Wang18p13350}%
  \BibitemOpen
  \bibfield  {author} {\bibinfo {author} {\bibfnamefont {Z.}~\bibnamefont
  {Wang}}, \bibinfo {author} {\bibfnamefont {D.}~\bibnamefont {Danovich}},
  \bibinfo {author} {\bibfnamefont {R.}~\bibnamefont {Ramanan}},\ and\ \bibinfo
  {author} {\bibfnamefont {S.}~\bibnamefont {Shaik}},\ }\bibfield  {title}
  {\bibinfo {title} {Oriented-external electric fields create absolute
  enantioselectivity in diels–alder reactions \: Importance of the molecular
  dipole moment},\ }\href@noop {} {\bibfield  {journal} {\bibinfo  {journal}
  {J. Am. Chem. Soc.}\ }\textbf {\bibinfo {volume} {140}},\ \bibinfo {pages}
  {13350} (\bibinfo {year} {2018})}\BibitemShut {NoStop}%
\bibitem [{\citenamefont {Olah}(2005)}]{olah05p2636}%
  \BibitemOpen
  \bibfield  {author} {\bibinfo {author} {\bibfnamefont {G.~A.}\ \bibnamefont
  {Olah}},\ }\bibfield  {title} {\bibinfo {title} {Beyond oil and gas: The
  methanol economy},\ }\href@noop {} {\bibfield  {journal} {\bibinfo  {journal}
  {Angew. Chem. Int. Ed.}\ }\textbf {\bibinfo {volume} {44}},\ \bibinfo {pages}
  {2636} (\bibinfo {year} {2005})}\BibitemShut {NoStop}%
\bibitem [{\citenamefont {Ke}\ \emph {et~al.}(2021)\citenamefont {Ke},
  \citenamefont {He}, \citenamefont {Liu}, \citenamefont {Ru}, \citenamefont
  {Liu},\ and\ \citenamefont {Lin}}]{Ke21p10860}%
  \BibitemOpen
  \bibfield  {author} {\bibinfo {author} {\bibfnamefont {C.}~\bibnamefont
  {Ke}}, \bibinfo {author} {\bibfnamefont {W.}~\bibnamefont {He}}, \bibinfo
  {author} {\bibfnamefont {S.}~\bibnamefont {Liu}}, \bibinfo {author}
  {\bibfnamefont {X.}~\bibnamefont {Ru}}, \bibinfo {author} {\bibfnamefont
  {S.}~\bibnamefont {Liu}},\ and\ \bibinfo {author} {\bibfnamefont
  {Z.}~\bibnamefont {Lin}},\ }\bibfield  {title} {\bibinfo {title} {Multiscale
  catalyst design for steam methane reforming assisted by deep learning},\
  }\href@noop {} {\bibfield  {journal} {\bibinfo  {journal} {J. Phys. Chem. C}\
  }\textbf {\bibinfo {volume} {125}},\ \bibinfo {pages} {10860} (\bibinfo
  {year} {2021})}\BibitemShut {NoStop}%
\bibitem [{\citenamefont {Han}\ \emph {et~al.}(2021)\citenamefont {Han},
  \citenamefont {Sarker}, \citenamefont {Ouyang}, \citenamefont {Mazheika},
  \citenamefont {Gao},\ and\ \citenamefont {Levchenko}}]{han21p1833}%
  \BibitemOpen
  \bibfield  {author} {\bibinfo {author} {\bibfnamefont {Z.-K.}\ \bibnamefont
  {Han}}, \bibinfo {author} {\bibfnamefont {D.}~\bibnamefont {Sarker}},
  \bibinfo {author} {\bibfnamefont {R.}~\bibnamefont {Ouyang}}, \bibinfo
  {author} {\bibfnamefont {A.}~\bibnamefont {Mazheika}}, \bibinfo {author}
  {\bibfnamefont {Y.}~\bibnamefont {Gao}},\ and\ \bibinfo {author}
  {\bibfnamefont {S.~V.}\ \bibnamefont {Levchenko}},\ }\bibfield  {title}
  {\bibinfo {title} {Single-atom alloy catalysts designed by first-principles
  calculations and artificial intelligence},\ }\href@noop {} {\bibfield
  {journal} {\bibinfo  {journal} {Nat. Commun.}\ }\textbf {\bibinfo {volume}
  {12}},\ \bibinfo {pages} {1833} (\bibinfo {year} {2021})}\BibitemShut
  {NoStop}%
\bibitem [{\citenamefont {Fu}\ \emph {et~al.}(2020)\citenamefont {Fu},
  \citenamefont {Yang},\ and\ \citenamefont {Wu}}]{fu2020p156001}%
  \BibitemOpen
  \bibfield  {author} {\bibinfo {author} {\bibfnamefont {Z.}~\bibnamefont
  {Fu}}, \bibinfo {author} {\bibfnamefont {B.}~\bibnamefont {Yang}},\ and\
  \bibinfo {author} {\bibfnamefont {R.}~\bibnamefont {Wu}},\ }\bibfield
  {title} {\bibinfo {title} {Understanding the activity of single-atom
  catalysis from frontier orbitals},\ }\href@noop {} {\bibfield  {journal}
  {\bibinfo  {journal} {Phys. Rev. Lett.}\ }\textbf {\bibinfo {volume} {125}},\
  \bibinfo {pages} {156001} (\bibinfo {year} {2020})}\BibitemShut {NoStop}%
\bibitem [{\citenamefont {Chen}\ \emph {et~al.}(2016)\citenamefont {Chen},
  \citenamefont {Urushihara}, \citenamefont {Chan},\ and\ \citenamefont
  {Nørskov}}]{Chen16p7133}%
  \BibitemOpen
  \bibfield  {author} {\bibinfo {author} {\bibfnamefont {L.~D.}\ \bibnamefont
  {Chen}}, \bibinfo {author} {\bibfnamefont {M.}~\bibnamefont {Urushihara}},
  \bibinfo {author} {\bibfnamefont {K.}~\bibnamefont {Chan}},\ and\ \bibinfo
  {author} {\bibfnamefont {J.~K.}\ \bibnamefont {Nørskov}},\ }\bibfield
  {title} {\bibinfo {title} {Electric field effects in electrochemical co$_2$
  reduction},\ }\href {https://doi.org/10.1021/acscatal.6b02299} {\bibfield
  {journal} {\bibinfo  {journal} {ACS Catal.}\ }\textbf {\bibinfo {volume}
  {6}},\ \bibinfo {pages} {7133} (\bibinfo {year} {2016})}\BibitemShut
  {NoStop}%
\bibitem [{\citenamefont {Michaelides}\ \emph {et~al.}(2003)\citenamefont
  {Michaelides}, \citenamefont {Liu}, \citenamefont {Zhang}, \citenamefont
  {Alavi}, \citenamefont {King},\ and\ \citenamefont
  {Hu}}]{Michaelides03p3704}%
  \BibitemOpen
  \bibfield  {author} {\bibinfo {author} {\bibfnamefont {A.}~\bibnamefont
  {Michaelides}}, \bibinfo {author} {\bibfnamefont {Z.-P.}\ \bibnamefont
  {Liu}}, \bibinfo {author} {\bibfnamefont {C.~J.}\ \bibnamefont {Zhang}},
  \bibinfo {author} {\bibfnamefont {A.}~\bibnamefont {Alavi}}, \bibinfo
  {author} {\bibfnamefont {D.~A.}\ \bibnamefont {King}},\ and\ \bibinfo
  {author} {\bibfnamefont {P.}~\bibnamefont {Hu}},\ }\bibfield  {title}
  {\bibinfo {title} {Identification of general linear relationships between
  activation energies and enthalpy changes for dissociation reactions at
  surfaces},\ }\href@noop {} {\bibfield  {journal} {\bibinfo  {journal} {J. Am.
  Chem. Soc.}\ }\textbf {\bibinfo {volume} {125}},\ \bibinfo {pages} {3704}
  (\bibinfo {year} {2003})}\BibitemShut {NoStop}%
\bibitem [{\citenamefont {Abild-Pedersen}\ \emph {et~al.}(2007)\citenamefont
  {Abild-Pedersen}, \citenamefont {Greeley}, \citenamefont {Studt},
  \citenamefont {Rossmeisl}, \citenamefont {Munter}, \citenamefont {Moses},
  \citenamefont {Skúlason}, \citenamefont {Bligaard},\ and\ \citenamefont
  {Nørskov}}]{abild-pedersen07p016105}%
  \BibitemOpen
  \bibfield  {author} {\bibinfo {author} {\bibfnamefont {F.}~\bibnamefont
  {Abild-Pedersen}}, \bibinfo {author} {\bibfnamefont {J.}~\bibnamefont
  {Greeley}}, \bibinfo {author} {\bibfnamefont {F.}~\bibnamefont {Studt}},
  \bibinfo {author} {\bibfnamefont {J.}~\bibnamefont {Rossmeisl}}, \bibinfo
  {author} {\bibfnamefont {T.~R.}\ \bibnamefont {Munter}}, \bibinfo {author}
  {\bibfnamefont {P.~G.}\ \bibnamefont {Moses}}, \bibinfo {author}
  {\bibfnamefont {E.}~\bibnamefont {Skúlason}}, \bibinfo {author}
  {\bibfnamefont {T.}~\bibnamefont {Bligaard}},\ and\ \bibinfo {author}
  {\bibfnamefont {J.~K.}\ \bibnamefont {Nørskov}},\ }\bibfield  {title}
  {\bibinfo {title} {Scaling properties of adsorption energies for
  hydrogen-containing molecules on transition-metal surfaces},\ }\href@noop {}
  {\bibfield  {journal} {\bibinfo  {journal} {Phys. Rev. Lett.}\ }\textbf
  {\bibinfo {volume} {99}},\ \bibinfo {pages} {016105} (\bibinfo {year}
  {2007})}\BibitemShut {NoStop}%
\bibitem [{\citenamefont {Kresse}\ and\ \citenamefont
  {J}(1996{\natexlab{a}})}]{Kresse96p11169}%
  \BibitemOpen
  \bibfield  {author} {\bibinfo {author} {\bibfnamefont {G.}~\bibnamefont
  {Kresse}}\ and\ \bibinfo {author} {\bibfnamefont {F.}~\bibnamefont {J}},\
  }\bibfield  {title} {\bibinfo {title} {Efficient iterative schemes for ab
  initio total-energy calculations using a plane-wave basis set},\ }\href@noop
  {} {\bibfield  {journal} {\bibinfo  {journal} {Phys. Rev. B}\ }\textbf
  {\bibinfo {volume} {54}},\ \bibinfo {pages} {11169} (\bibinfo {year}
  {1996}{\natexlab{a}})}\BibitemShut {NoStop}%
\bibitem [{\citenamefont {Kresse}\ and\ \citenamefont
  {J}(1996{\natexlab{b}})}]{Kresse96p15}%
  \BibitemOpen
  \bibfield  {author} {\bibinfo {author} {\bibfnamefont {G.}~\bibnamefont
  {Kresse}}\ and\ \bibinfo {author} {\bibfnamefont {F.}~\bibnamefont {J}},\
  }\bibfield  {title} {\bibinfo {title} {Efficiency of ab-initio total energy
  calculations for metals and semiconductors using a plane-wave basis set},\
  }\href@noop {} {\bibfield  {journal} {\bibinfo  {journal} {Comput. Mater.
  Sci.}\ }\textbf {\bibinfo {volume} {6}},\ \bibinfo {pages} {15} (\bibinfo
  {year} {1996}{\natexlab{b}})}\BibitemShut {NoStop}%
\bibitem [{\citenamefont {Blochl}(1994)}]{Blochl94p17953}%
  \BibitemOpen
  \bibfield  {author} {\bibinfo {author} {\bibfnamefont {P.~E.}\ \bibnamefont
  {Blochl}},\ }\bibfield  {title} {\bibinfo {title} {Projector augmented-wave
  method},\ }\href@noop {} {\bibfield  {journal} {\bibinfo  {journal} {Phys.
  Rev. B}\ }\textbf {\bibinfo {volume} {50}},\ \bibinfo {pages} {17953}
  (\bibinfo {year} {1994})}\BibitemShut {NoStop}%
\bibitem [{\citenamefont {Perdew}\ and\ \citenamefont
  {Wang}(1992)}]{Perdew92p13244}%
  \BibitemOpen
  \bibfield  {author} {\bibinfo {author} {\bibfnamefont {J.~P.}\ \bibnamefont
  {Perdew}}\ and\ \bibinfo {author} {\bibfnamefont {Y.}~\bibnamefont {Wang}},\
  }\bibfield  {title} {\bibinfo {title} {Accurate and simple analytic
  representation of the electron-gas correlation energy},\ }\href@noop {}
  {\bibfield  {journal} {\bibinfo  {journal} {Phys. Rev. B}\ }\textbf {\bibinfo
  {volume} {45}},\ \bibinfo {pages} {13244} (\bibinfo {year}
  {1992})}\BibitemShut {NoStop}%
\bibitem [{\citenamefont {Neugebauer}\ and\ \citenamefont
  {Scheffler}(1992)}]{Neugebauer92p16067}%
  \BibitemOpen
  \bibfield  {author} {\bibinfo {author} {\bibfnamefont {J.}~\bibnamefont
  {Neugebauer}}\ and\ \bibinfo {author} {\bibfnamefont {M.}~\bibnamefont
  {Scheffler}},\ }\bibfield  {title} {\bibinfo {title} {Adsorbate-substrate and
  adsorbate-adsorbate interactions of \text{Na} and \text{K} adlayers on
  \text{Al}(111)},\ }\href {https://doi.org/10.1103/PhysRevB.46.16067}
  {\bibfield  {journal} {\bibinfo  {journal} {Phys. Rev. B}\ }\textbf {\bibinfo
  {volume} {46}},\ \bibinfo {pages} {16067} (\bibinfo {year}
  {1992})}\BibitemShut {NoStop}%
\bibitem [{\citenamefont {Henkelman}\ \emph {et~al.}(2000)\citenamefont
  {Henkelman}, \citenamefont {Uberuaga},\ and\ \citenamefont
  {J{\'{o}}nsson}}]{Henkelman00p9901}%
  \BibitemOpen
  \bibfield  {author} {\bibinfo {author} {\bibfnamefont {G.}~\bibnamefont
  {Henkelman}}, \bibinfo {author} {\bibfnamefont {B.~P.}\ \bibnamefont
  {Uberuaga}},\ and\ \bibinfo {author} {\bibfnamefont {H.}~\bibnamefont
  {J{\'{o}}nsson}},\ }\bibfield  {title} {\bibinfo {title} {A climbing image
  nudged elastic band method for finding saddle points and minimum energy
  paths},\ }\href {https://doi.org/10.1063/1.1329672} {\bibfield  {journal}
  {\bibinfo  {journal} {J. Chem. Phys.}\ }\textbf {\bibinfo {volume} {113}},\
  \bibinfo {pages} {9901} (\bibinfo {year} {2000})}\BibitemShut {NoStop}%
\bibitem [{\citenamefont {Henkelman}\ and\ \citenamefont
  {Jónsson}(1999)}]{Henkelman99p7010}%
  \BibitemOpen
  \bibfield  {author} {\bibinfo {author} {\bibfnamefont {G.}~\bibnamefont
  {Henkelman}}\ and\ \bibinfo {author} {\bibfnamefont {H.}~\bibnamefont
  {Jónsson}},\ }\bibfield  {title} {\bibinfo {title} {A dimer method for
  finding saddle points on high dimensional potential surfaces using only first
  derivatives},\ }\href {https://doi.org/10.1063/1.480097} {\bibfield
  {journal} {\bibinfo  {journal} {J. Chem. Phys.}\ }\textbf {\bibinfo {volume}
  {111}},\ \bibinfo {pages} {7010} (\bibinfo {year} {1999})}\BibitemShut
  {NoStop}%
\bibitem [{\citenamefont {Yeh}\ \emph {et~al.}(2019)\citenamefont {Yeh},
  \citenamefont {Pham}, \citenamefont {Nachimuthu},\ and\ \citenamefont
  {Jiang}}]{Yeh19p8230}%
  \BibitemOpen
  \bibfield  {author} {\bibinfo {author} {\bibfnamefont {C.-H.}\ \bibnamefont
  {Yeh}}, \bibinfo {author} {\bibfnamefont {T.~M.~L.}\ \bibnamefont {Pham}},
  \bibinfo {author} {\bibfnamefont {S.}~\bibnamefont {Nachimuthu}},\ and\
  \bibinfo {author} {\bibfnamefont {J.-C.}\ \bibnamefont {Jiang}},\ }\bibfield
  {title} {\bibinfo {title} {Effect of external electric field on methane
  conversion on iro$_2$ (110) surface: A density functional theory study},\
  }\href {https://doi.org/10.1021/acscatal.9b01910} {\bibfield  {journal}
  {\bibinfo  {journal} {ACS Catal.}\ }\textbf {\bibinfo {volume} {9}},\
  \bibinfo {pages} {8230} (\bibinfo {year} {2019})}\BibitemShut {NoStop}%
\bibitem [{\citenamefont {Pacchioni}\ \emph {et~al.}(1997)\citenamefont
  {Pacchioni}, \citenamefont {Lomas},\ and\ \citenamefont
  {Illas}}]{Pacchioni97p263}%
  \BibitemOpen
  \bibfield  {author} {\bibinfo {author} {\bibfnamefont {G.}~\bibnamefont
  {Pacchioni}}, \bibinfo {author} {\bibfnamefont {J.~R.}\ \bibnamefont
  {Lomas}},\ and\ \bibinfo {author} {\bibfnamefont {F.}~\bibnamefont {Illas}},\
  }\bibfield  {title} {\bibinfo {title} {Electric field effects in
  heterogeneous catalysis},\ }\href
  {https://doi.org/10.1016/S1381-1169(96)00490-6} {\bibfield  {journal}
  {\bibinfo  {journal} {J. Mol. Catal. A}\ }\textbf {\bibinfo {volume} {119}},\
  \bibinfo {pages} {263} (\bibinfo {year} {1997})}\BibitemShut {NoStop}%
\bibitem [{\citenamefont {Che}\ \emph {et~al.}(2016)\citenamefont {Che},
  \citenamefont {Ha},\ and\ \citenamefont {{McEwen}}}]{Che16p77}%
  \BibitemOpen
  \bibfield  {author} {\bibinfo {author} {\bibfnamefont {F.}~\bibnamefont
  {Che}}, \bibinfo {author} {\bibfnamefont {S.}~\bibnamefont {Ha}},\ and\
  \bibinfo {author} {\bibfnamefont {J.-S.}\ \bibnamefont {{McEwen}}},\
  }\bibfield  {title} {\bibinfo {title} {Elucidating the field influence on the
  energetics of the methane steam reforming reaction: A density functional
  theory study},\ }\href {https://doi.org/10.1016/j.apcatb.2016.04.026}
  {\bibfield  {journal} {\bibinfo  {journal} {Appl. Catal. B}\ }\textbf
  {\bibinfo {volume} {195}},\ \bibinfo {pages} {77} (\bibinfo {year}
  {2016})}\BibitemShut {NoStop}%
\bibitem [{\citenamefont {Campbell}\ and\ \citenamefont
  {Mao}(2021)}]{Campbell21p647}%
  \BibitemOpen
  \bibfield  {author} {\bibinfo {author} {\bibfnamefont {C.~T.}\ \bibnamefont
  {Campbell}}\ and\ \bibinfo {author} {\bibfnamefont {Z.}~\bibnamefont {Mao}},\
  }\bibfield  {title} {\bibinfo {title} {Analysis and prediction of reaction
  kinetics using the degree of rate control},\ }\href@noop {} {\bibfield
  {journal} {\bibinfo  {journal} {J. Catal.}\ }\textbf {\bibinfo {volume}
  {404}},\ \bibinfo {pages} {647} (\bibinfo {year} {2021})}\BibitemShut
  {NoStop}%
\bibitem [{\citenamefont {Valdés-Solís}\ \emph {et~al.}(2006)\citenamefont
  {Valdés-Solís}, \citenamefont {Marbán},\ and\ \citenamefont
  {Fuertes}}]{Valdes06p354}%
  \BibitemOpen
  \bibfield  {author} {\bibinfo {author} {\bibfnamefont {T.}~\bibnamefont
  {Valdés-Solís}}, \bibinfo {author} {\bibfnamefont {G.}~\bibnamefont
  {Marbán}},\ and\ \bibinfo {author} {\bibfnamefont {A.}~\bibnamefont
  {Fuertes}},\ }\bibfield  {title} {\bibinfo {title} {Nanosized catalysts for
  the production of hydrogen by methanol steam reforming},\ }\href
  {https://doi.org/10.1016/j.cattod.2006.05.063} {\bibfield  {journal}
  {\bibinfo  {journal} {Catal. Today}\ }\textbf {\bibinfo {volume} {116}},\
  \bibinfo {pages} {354} (\bibinfo {year} {2006})}\BibitemShut {NoStop}%
\bibitem [{\citenamefont {Twigg}\ and\ \citenamefont
  {Spencer}(2003)}]{Twigg03p191}%
  \BibitemOpen
  \bibfield  {author} {\bibinfo {author} {\bibfnamefont {M.~V.}\ \bibnamefont
  {Twigg}}\ and\ \bibinfo {author} {\bibfnamefont {M.~S.}\ \bibnamefont
  {Spencer}},\ }\bibfield  {title} {\bibinfo {title} {Deactivation of copper
  metal catalysts for methanol decomposition, methanol steam reforming and
  methanol synthesis},\ }\href@noop {} {\bibfield  {journal} {\bibinfo
  {journal} {Top. Catal.}\ }\textbf {\bibinfo {volume} {22}},\ \bibinfo {pages}
  {191} (\bibinfo {year} {2003})}\BibitemShut {NoStop}%
\bibitem [{\citenamefont {Ke}\ and\ \citenamefont {Lin}(2020)}]{Ke20p349}%
  \BibitemOpen
  \bibfield  {author} {\bibinfo {author} {\bibfnamefont {C.}~\bibnamefont
  {Ke}}\ and\ \bibinfo {author} {\bibfnamefont {Z.}~\bibnamefont {Lin}},\
  }\bibfield  {title} {\bibinfo {title} {Density functional theory based micro-
  and macro-kinetic studies of ni-catalyzed methanol steam reforming},\ }\href
  {https://doi.org/10.3390/catal10030349} {\bibfield  {journal} {\bibinfo
  {journal} {Catalysts}\ }\textbf {\bibinfo {volume} {10}},\ \bibinfo {pages}
  {349} (\bibinfo {year} {2020})}\BibitemShut {NoStop}%
\end{thebibliography}%
\end{document}